\documentclass[a4paper]{jpconf}
\usepackage{graphicx}

\usepackage{amssymb}
\usepackage{amsmath}
\usepackage{graphicx}
\usepackage[normalem]{ulem}
\usepackage[dvips]{color}
\renewcommand\sout{\bgroup \color{red} \ULdepth=-.5ex \ULset}

\begin{document}
\title{Imprints of Nuclear Symmetry Energy on Properties of Neutron Stars}

\author{Bao-An Li$^1$, Lie-Wen Chen$^{1,2}$, Michael Gearheart$^1$, Joshua Hooker$^1$, Che Ming Ko$^3$, Plamen G. Krastev$^{1,4}$, Wei-Kang Lin$^1$,
William G. Newton$^1$, De-Hua Wen$^{1,5}$, Chang Xu$^{1,6}$ and
Jun Xu$^3$}
\address{$^1$Department of Physics and Astronomy, Texas A$\&$M
University-Commerce, Commerce, Texas 75429-3011, USA}
\address{$^2$Department of Physics, Shanghai Jiao Tong University,
Shanghai 200240, China}
\address{$^3$Cyclotron Institute and Department of Physics and Astronomy,
Texas A$\&$M University, College Station, Texas 77843-3366, USA}
\address{$^4$Department of Physics, San Diego State University,
5500 Campanile Drive, San Diego, CA 92182-1233, U.S.A.}
\address{$^5$Department of Physics, South China University of
Technology, Guangzhou 510641, China}
\address{$^6$Department of Physics, Nanjing
University, Nanjing 210008, China}

\ead{Bao-An$\_$Li@Tamu-Commerce.edu}

\begin{abstract}
Significant progress has been made in recent years in constraining
the density dependence of nuclear symmetry energy using
terrestrial nuclear laboratory data. Around and below the nuclear
matter saturation density, the experimental constraints start to
merge in a relatively narrow region. At supra-saturation
densities, there are, however, still large uncertainties. After
summarizing the latest experimental constraints on the density
dependence of nuclear symmetry energy, we highlight a few recent
studies examining imprints of nuclear symmetry energy on the
binding energy, energy release during hadron-quark phase
transitions as well as the $w$-mode frequency and damping time of
gravitational wave emission of neutron stars.
\end{abstract}

\section{Introduction}

Properties of neutron-rich matter are at the heart of many
fundamental questions in both nuclear physics and astrophysics
\cite{Chuck10}. They are currently being explored experimentally
by using a wide variety of advanced new facilities, such as,
Facility for Rare Isotope Beams (FRIB), x-ray satellites and
gravitational wave detectors. Most critical to understanding
experimental observations of various phenomena using these
facilities is the Equation of State (EOS) of neutron-rich
nucleonic matter. The latter can be written in terms of the energy
per nucleon at density $\rho$ as
\begin{equation} E(\rho ,\delta
)=E(\rho ,\delta =0)+E_{\rm sym}(\rho )\delta
^{2}+\mathcal{O}(\delta^4)
\end{equation}
where $\delta\equiv(\rho_{n}-\rho _{p})/(\rho _{p}+\rho _{n})$ is
the neutron-proton asymmetry and $E_{\rm sym}(\rho)$ is the
density-dependent nuclear symmetry energy. The latter is a vital
ingredient in the theoretical description of neutron stars where
the proton fraction is uniquely determined by the symmetry energy
$E_{\rm sym}(\rho)$ \cite{Lat00,steiner05}, and of the structure
of neutron-rich nuclei and reactions involving them
\cite{ireview98,ibook01,ditoro,Chen07r,LCK08}. It is currently the
most uncertain part of the EOS of neutron-rich nuclear matter,
especially at super-saturation densities. Very recently, attempts
have been made to extract meaningful constraints on the neutron
star EOS, facilitated by neutron star observational data with
sufficient accuracy, and from a wide enough range of neutron star
phenomena associated with different density regions within the
star \cite{Read09,Ozel10,Steiner10}. This signals a new era in
which the quality and quantity of astronomical data make
constraining the microphysical properties of neutron star matter
from sub-nuclear to super-nuclear densities a realistic endeavor.
At the same time, the nuclear matter EOS is being probed with
unprecedented accuracy by a variety of terrestrial nuclear
experiments, see, e.g., refs.\
\cite{LCK08,dan02,Fuc06Kaon,GSI,MSU,RIKEN,Umesh}. Combining the
observational and experimental data, together with the
astrophysical and nuclear modeling required to interpret them, has
the potential to impose stringent constraints on the nuclear
symmetry energy at both sub- and super-saturation densities, see,
e.g., refs.\ \cite{Ste05b,LiBA06a,Aaron,Plamen08,XCLM09,Newton}.

\section{The key physics behind nuclear symmetry energy}

Why is the nuclear symmetry energy so uncertain? To help
understand this question, it is useful to first recall the
relationship between the symmetry energy and the
isospin-dependence of strong interaction at the mean-field level.
According to the well-known Lane potential \cite{Lan62}, the
neutron/proton single-particle potential $U_{n/p}(\rho,k,\delta)$
can be well approximated by $ U_{n/p}(\rho,k,\delta)\approx
U_0(\rho,k) \pm U_{\rm sym}(\rho,k)\delta $ where the
$U_0(\rho,k)$ and $U_{\rm sym}(\rho,k)$ are, respectively, the
isoscalar and isovector (symmetry) potentials for nucleons with
momentum $k$ in nuclear matter of isospin asymmetry $\delta$ at
density $\rho$. It was shown first by Brueckner, Dabrowski and
Haensel \cite{bru64,Dab73} using K-matrices within the Brueckner
theory in the 60's-70's, and more recently by Xu et al.
\cite{XuLiChen10a} using the Hugenholtz-Van Hove (HVH) theorem
\cite{hug} that the nuclear symmetry energy can be explicitly
expressed as
\begin{equation}
E_{\rm sym}(\rho) = \frac{1}{6} \frac{\partial(t+U_0)}{\partial
k}\mid _{k_F}\cdot k_F + \frac{1}{2}U_{\rm sym}(\rho,k_F),
\end{equation}
where $t(k)=\hbar ^2 k^2 / 2m$ is the kinetic energy and
$k_F=(3\pi^2\rho/2)^{1/3}$ is the nucleon Fermi momentum in
symmetric nuclear matter at density $\rho$. Moreover, it was shown
recently using the HVH theorem that the slope of nuclear symmetry
energy at an arbitrary density $\rho$ can be written as
\cite{XuLiChen10a,xuli2}
\begin{equation}
L(\rho)\equiv 3
\rho \frac{\partial E_{\rm sym}(\rho)}{\partial \rho}|_{\rho}=
\frac{1}{6} \frac{\partial (t+U_0)}{\partial k}|_{k_F} \cdot k_F +
\frac{1}{6} \frac{\partial^2 (t+U_0)}{\partial k^2}|_{k_F} \cdot
 k_F^2 + \frac{3}{2} U_{\rm sym}(\rho,k_F)+ \frac{\partial U_{\rm sym}}{\partial k}|_{k_F} \cdot k_F.
\end{equation}
The $E_{sym}(\rho)$ and $L(\rho)$ are correlated by the same
underlying interaction. The above expressions for $E_{sym}(\rho)$
and $L(\rho)$ in terms of the isoscalar and isovector
single-particle potentials are particularly useful for extracting
the symmetry energy and its density slope from terrestrial nuclear
laboratory experiments. On one hand, in many reaction models, such
as transport model simulations of nuclear reactions, the EOS only
enters indirectly into the reaction dynamics and affects the final
observables through the single-nucleon potential
$U_{n/p}(\rho,k,\delta)$. Except in situations where statistical
equilibrium is established and thus many observables are directly
related to the energy $E(\rho,\delta)$ after correcting for
finite-size effects, what is being directly probed in nuclear
reactions is the single-nucleon potential
$U_{n/p}(\rho,k,\delta)$. On the other hand, in many structure
models, such as shell models, the direct input is also the single
nucleon potential. Thus, the isospin-momentum-density dependence
of the single nucleon potential $U_{n/p}(\rho,k,\delta)$ is the
key in determining the EOS of neutron-rich nuclear matter.
Comparing model calculations with experimental data allows the
determination of the single nucleon potential which then
determines directly the $E_{sym}(\rho)$ and $L(\rho)$. While the
density and momentum dependence of the isoscalar potential
$U_{0}(\rho,k)$ has been relatively well determined up to about 4
to 5 times the normal nuclear matter density $\rho_0$ using
nucleon global optical potentials from nucleon-nucleus scatterings
as well as kaon production and nuclear collective flow in
relativistic heavy-ion collisions \cite{dan02,Fuc06Kaon}, our
current knowledge about the symmetry potential $U_{\rm
sym}(\rho,k)$ is rather poor especially at high density and/or
momenta \cite{LCK08,Fuchs05,zuo05,ria05,ria06,Gio10}. In fact,
while some models predict decreasing symmetry potentials albeit at
different rates, some others predict instead increasing ones with
growing nucleon momentum. Thus, the high density/mometum behavior
of the symmetry potential $U_{\rm sym}(\rho,k)$ is still very
uncertain. Experimentally, there is some constraints on the
symmetry potential only at normal density for low energy nucleons
up to about 100 MeV obtained from nucleon-nucleus and (p, n)
charge exchange reactions \cite{XuLiChen10a}.

Of course, besides the uncertain density and momentum dependence
of the symmetry potential, different approaches used in treating
many-body problems in the various models also contribute to the
divergence of the predicted symmetry energy especially at
supra-saturation densities. Nonetheless, there are several key and
commonly used physics ingredients that can affect the predicted
$E_{\rm sym}(\rho)$ in all theories. In the interacting Fermi gas
model, the isovector potential at $k_F$ can be written as
\cite{pre,Xu-tensor}
\begin{equation}
U_{\rm sym}(k_F,\rho)=
\frac{1}{4}\rho\int
[V_{T1}(r_{ij})f^{T1}(r_{ij})-V_{T0}(r_{ij})f^{T0}(r_{ij})]d^3r_{ij}
\end{equation}
in terms of the isosinglet (T=0) and isotriplet (T=1)
nucleon-nucleon (NN) interactions $V_{T0}(r_{ij})$ and
$V_{T1}(r_{ij})$, and the corresponding NN correlation functions
$f^{T0}(r_{ij})$ and $f^{T1}(r_{ij})$, respectively. Needless to
say, if there is no isospin dependence in both the NN interaction
and correlation function, then the isovector potential $ U_{\rm
sym}(k_F,\rho)$ vanishes. The $E_{\rm sym}(\rho)$ thus reflects
the competition between the NN interaction strengths and
correlation functions of the isosinglet and isotriplet channels.
However, effects of several key ingredients, such as the tensor
force, isospin-dependence of short-range nucleon-nucleon
correlations and spin-isospin dependent three-body force, are not
yet well understood, see, e.g., refs.\
\cite{Xu-tensor,Lee11,Dong09,Fri05}. Consequently, the predicted
symmetry energies diverge rather widely, especially at
supra-saturation densities.

\section{Current status of experimental constraints on nuclear symmetry energy}
\begin{figure}[htb]
\begin{center}
\begin{minipage}{13pc}
\includegraphics[scale=0.7]{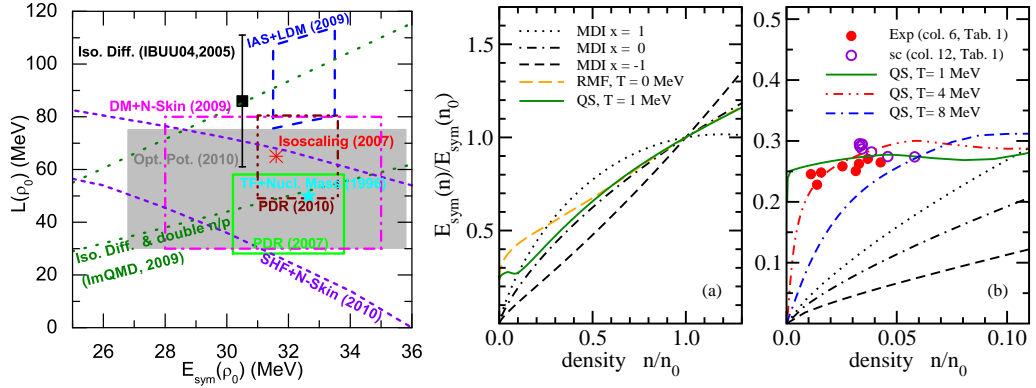}
\end{minipage}
\begin{minipage}{14pc}
\includegraphics[scale=0.4]{Joe.eps}
\end{minipage}\label{sl-plot}
\caption{{\protect\small \textbf{Left Window}: Density slope
versus the magnitude of the symmetry energy at normal density
extracted from isospin diffusion and neutron/proton ratio of
pre-equilibrium nucleon emissions within the Improved Molecular
Dynamics (ImQMD-2009) model \cite{tsa1,tsa2,Fami}, isospin
diffusion within the Isospin-Dependent Boltzmann-Uehling-Ulenbeck
(IBUU04- 2005) \cite{chen05a,li05a}, isoscaling \cite{shet},
energy shift of isobaric analogue states within liquid drop model
(IAS+LDM-2009) \cite{Pawel09}, neutron-skins of several heavy
nuclei using the droplet model (DM) or the Skyrme-Hartree-Fock
(SHF) approach \cite{Cen09,War09,Chen10}, pygmy dipole resonances
(PDR) in $^{209}$Pb, $^{68}$Ni and $^{132}$Sn \cite{Kli07,Car10},
and the nucleon global optical potentials
(GOP)~\cite{XuLiChen10a}. Taken from ref. \cite{XuLiChen10a}.
\textbf{Right Window}: Comparisons of the scaled symmetry energy
as a function of the scaled total density $n/n_0$ for different
approaches and the experiment. Left panel: Symmetry energies from
the MDI parametrization of Chen et al. \cite{chen05a,Che07} for
$T=0$ and different $L(\rho_0)$, controlled by the parameter $x$
(dotted, dot-dashed and dashed (black) lines); from the
quantum-statistical (QS) approach including light clusters for
temperature $T=1$~MeV (solid (green) line), and from the
relativistic mean-field (RMF) model at $T=0$ including heavy
clusters (long-dashed (orange) line). Right panel: The internal
scaled symmetry energy in an expanded low-density region. Shown
are again the MDI curves and the QS results for $T=1,4,$ and 8~MeV
compared to the experimental data using the entropy from the
nuclear statistical equilibrium (NSE) model (solid circles) and
the results of the self-consistent calculation (open circles).
Taken from ref.~\cite{Joe10}.}}
\end{center}
\end{figure}
While significant progress has been made in constraining the
$E_{\rm sym}(\rho)$ as a result of the great efforts made by many
people in both the nuclear physics and the astrophysics community,
experimental constraints obtained so far are still not tight
enough to limit stringently model predictions even around the
saturation density $\rho_0$. Moreover, it has been broadly
recognized that significant model dependence exists for most of
the observables studied so far. For instance, near $\rho_0$ the
symmetry energy can be characterized by using the value of $E_{\rm
sym}(\rho_0)$ and the slope parameter $L(\rho_0)$. The $E_{\rm
sym}(\rho_0)$ is known to be around $28-34$ MeV mainly from
analyzing nuclear masses within liquid-drop models
\cite{mye63,Peter95,pomo}. The exact value of $E_{\rm
sym}(\rho_0)$ extracted this way depends largely on the size and
accuracy of nuclear mass data base as well as whether/what surface
symmetry energy is used in the analysis. The latest study by Liu
et al. \cite{Mliu10} by analyzing the atomic masses indicates that
the value of $E_{\rm sym}(\rho_0)$ is between $29.4$ and $32.8$
MeV while the value of $L(\rho_0)$ is between 52.9 and 78.7 MeV.
The density slope $L(\rho_0)$ has also been extensively studied
but remains much more uncertain. Shown in the left window of Fig.\
1 are the $L(\rho_0)$ values extracted from studying several
phenomena/observables in nuclear structures and reactions
\cite{XuLiChen10a,tsa1,tsa2,Fami,chen05a,li05a,shet,Pawel09,Cen09,War09,Chen10,Kli07,Car10}.
The values of $L(\rho_0)$ from these studies scatter between about
20 to 115 MeV while each individual study may have given a smaller
uncertain range.   Although it is very encouraging to note that
the constraints seem to start converging around $L(\rho_0)\approx
60$ MeV and $E_{\rm sym}(\rho_0)\approx 31$ MeV, it is obviously
important to further narrow them down with new measurements and
model analyses. It is thus encouraging to see the most recent
determination of $E_{\rm sym}(\rho _{0})=$ $30.5\pm 3$ MeV and
$L(\rho_0)=52.5\pm 20$ MeV~\cite{Che11} from combining the
constraints on $L(\rho_0)$ and $E_{\rm sym}(\rho_0)$ based on
recently extracted $E_{\rm sym}(\rho_A = 0.1$
fm$^{-3})$~\cite{Mliu10,Tri08,Cao08} with those from recent
analyses~\cite{Chen10} of neutron skin thickness of Sn isotopes in
the same SHF approach. At very low densities, the formation of
clusters is energetically favored. The symmetry energy of the
clustered matter is expected to be different from that of uniform
matter, see, e.g., refs. \cite{Hor06,Sam09,Typ10}. Indeed, it was
recently found in experiments by Natowitz et al. that the symmetry
energy of clustered matter remains finite as the average density
approaches zero \cite{Joe10}. As shown in the right window of
Fig.\ 1, conventional theoretical calculations of the symmetry
energy of uniform matter based on mean-field approaches (e.g., the
MDI interaction\cite{das,chen05a,Che07}) fail to give the correct
low-temperature, low-density limit that is governed by
correlations, in particular by the appearance of bound states. A
recently developed quantum statistical (QS) approach \cite{Typ10}
that takes the formation of clusters into account predicts
symmetry energies that are in very good agreement with the
experimental data. At supra-saturation densities, the situation is
much worse. In fact, even the trend of the symmetry energy,
namely, whether it increases or decreases with increasing density,
is still controversial partially because of the very limited data
available and the less known clean probes compared to the
situation at sub-saturation densities. Among all observables
studied so far, the neutron-proton differential flow
\cite{LiBA00}, $\pi^-/\pi^+$ ratio \cite{LiBA02} and the
squeeze-out of the neutron/proton ratio \cite{Yong07} especially
at high transverse momenta in heavy-ion reactions are among the
most promising probes of the high density behavior of the nuclear
symmetry energy. Unfortunately, the conclusions are still model
dependent \cite{XiaoPRL,XuKo10,Feng10,Rus11,Coz11}.

\section{Impacts of nuclear symmetry energy on properties of neutron stars}

It is well known that many critical issues in astrophysics, such
as the composition and the cooling rate of proto-neutron stars as
well as the core-crust transition density, properties of the pasta
phase, the mass-radius correlation and the moment of inertia of
neutron stars, all depend sensitively on the $E_{\rm sym}(\rho)$.
Many research papers and review articles have been written on
these topics, see, e.g., refs. \cite{Lat00,steiner05,hor}. Most of
these studies have concentrated on astrophysical effects of the
symmetry energy at sub-saturation densities. However, the nuclear
symmetry energy at supra-saturation densities has long been
considered by some astrophysicists as the most uncertain one among
all properties of dense neutron-rich nucleonic matter
\cite{kut,ste06,kub07}. More efforts are thus need to explore many
possible astrophysical effects of the symmetry energy at
supra-saturation densities \cite{Wen09}. In this section, as
examples we highlight a few recent studies on effects of the
symmetry energy, especially its high-density behavior, on the
binding energy of neutron stars, total energy release during
mini-collapse triggered by hadron-quark phase transitions inside
neutron stars, and the $w$-mode of gravitational waves from
neutron stars.

\subsection{Effects of symmetry energy on the binding energy of neutron stars}

\begin{figure}[htb]
\begin{minipage}{18pc}
\includegraphics[scale=0.5]{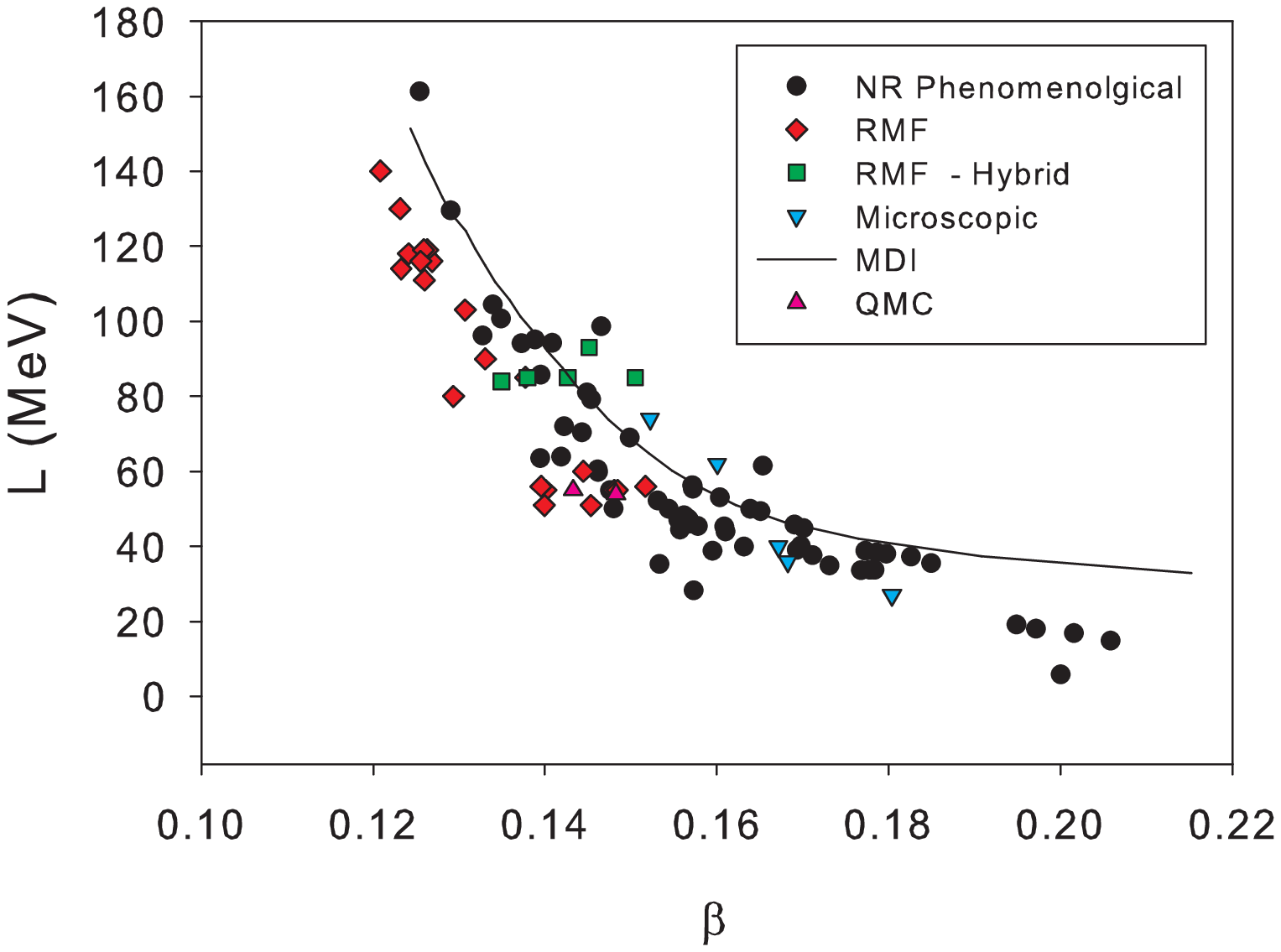}
\end{minipage}
\begin{minipage}{14pc}
\includegraphics[scale=0.5]{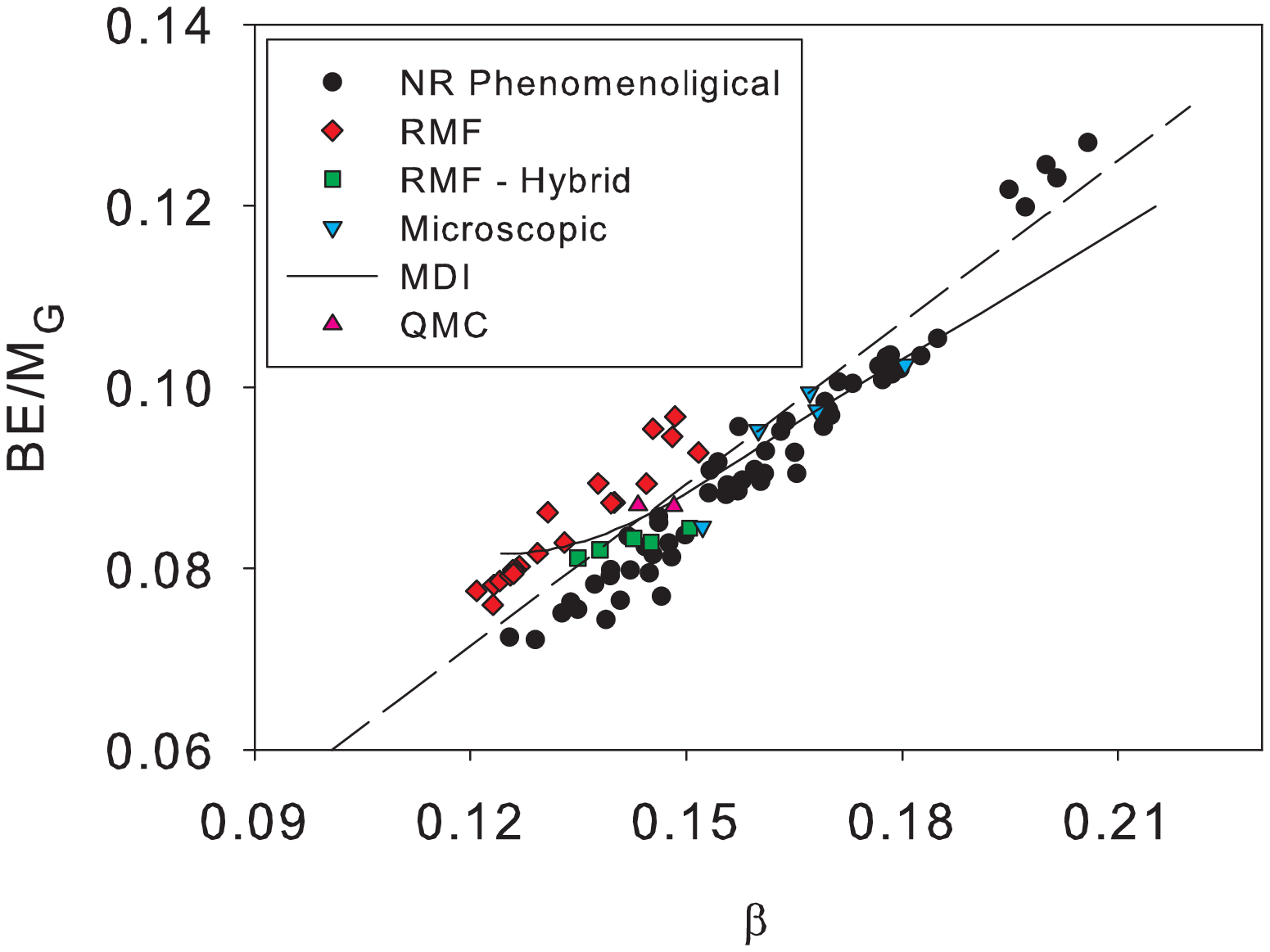}
\end{minipage}
\caption{\label{Newton1} {\protect\small \textbf{Left Window}:
Correlation between the slope parameter $L(\rho_0)$ of the
symmetry energy and the compactness parameter $\beta = G M_{\rm G}
/ R c^2$ for the sample of EoSs. \textbf{Right Window}: The
fractional gravitational binding energy of the neutron star
$BE/M_{\rm G}$ versus the compactness parameter $\beta$ for the
sample of EoSs. The dashed line indicates the analytic Newtonian
result.} Taken from ref. \cite{Newton}.}
\end{figure}

The binding energy of a neutron star is given by \cite{Weinberg} $
BE=M_G-M_{\rm bar}$, where $M_G$ is the gravitational mass of the
neutron star measured from infinity \cite{Gravity}, i.e.,
\begin{equation}
M_G=\int_0^R4\pi r^2\epsilon(r)dr
\end{equation}
with $\epsilon(r)$ being the energy density at radius r. The
$M_{\rm bar}$ is the baryonic mass of the neutron star, namely,
the mass when all the matter in the neutron star is dispersed to
infinity \cite{Weinberg}. It can be calculated from $M_{\rm
bar}=NM_B$, where $M_B=1.66\times10^{-24}g$ is the nucleon mass
and $N$ is the total number of baryons \cite{Weinberg}, i.e.,
\begin{equation}
N=\int^R_04\pi
r^2\left[1-\frac{2m(r)}{r}\right]^{-1/2}\rho_B(r)dr\,
\end{equation}
with $\rho_B(r)$ being the baryon density profile of the neutron
star. For a given EOS, the binding energy can be readily obtained
by first solving the Tolman-Opperheimer-Volkov (TOV) equation.
Newton et al. \cite{Newton} examined the correlation among the
symmetry energy slope $L(\rho_0)$, pressure at normal density
$P(n_0)$, compactness of neutron stars $\beta = G M_{\rm G} /
Rc^2$ and the binding energy. As shown in Figs.\ \ref{Newton1},
for a wide range of EoSs, the binding energy correlates roughly
linearly with the compactness of the neutron star $\beta$, which
in turn correlates strongly with the slope of the nuclear symmetry
energy $L(\rho_0)$ predicted by the EoSs. Therefore one expects to
see a correlation between $L(\rho_0)$ and the binding energy of
the neutron star, and consequently the baryon mass, given that the
gravitational mass is fixed. Thus, laboratory constraints on $L$
can be used to extract the constraint on the baryon mass of
neutron stars. This approach was used to constrain the lower mass
member of the double pulsar binary system, PSR J0737-3039B.
Comparing with independent constraints derived from modeling the
progenitor star of J0737-3039B up to and through its collapse
under the assumption that it was formed in an electron capture
supernova, it was found that the two sets of constraints are
consistent only if $L(\rho_0)\leq$ 70 MeV \cite{Newton}.

\subsection{Effects of symmetry energy on the energy release during hadron-quark phase transition in
neutron stars}

\begin{figure}[h]
\begin{minipage}{14pc}
\includegraphics[width=0.91\textwidth]{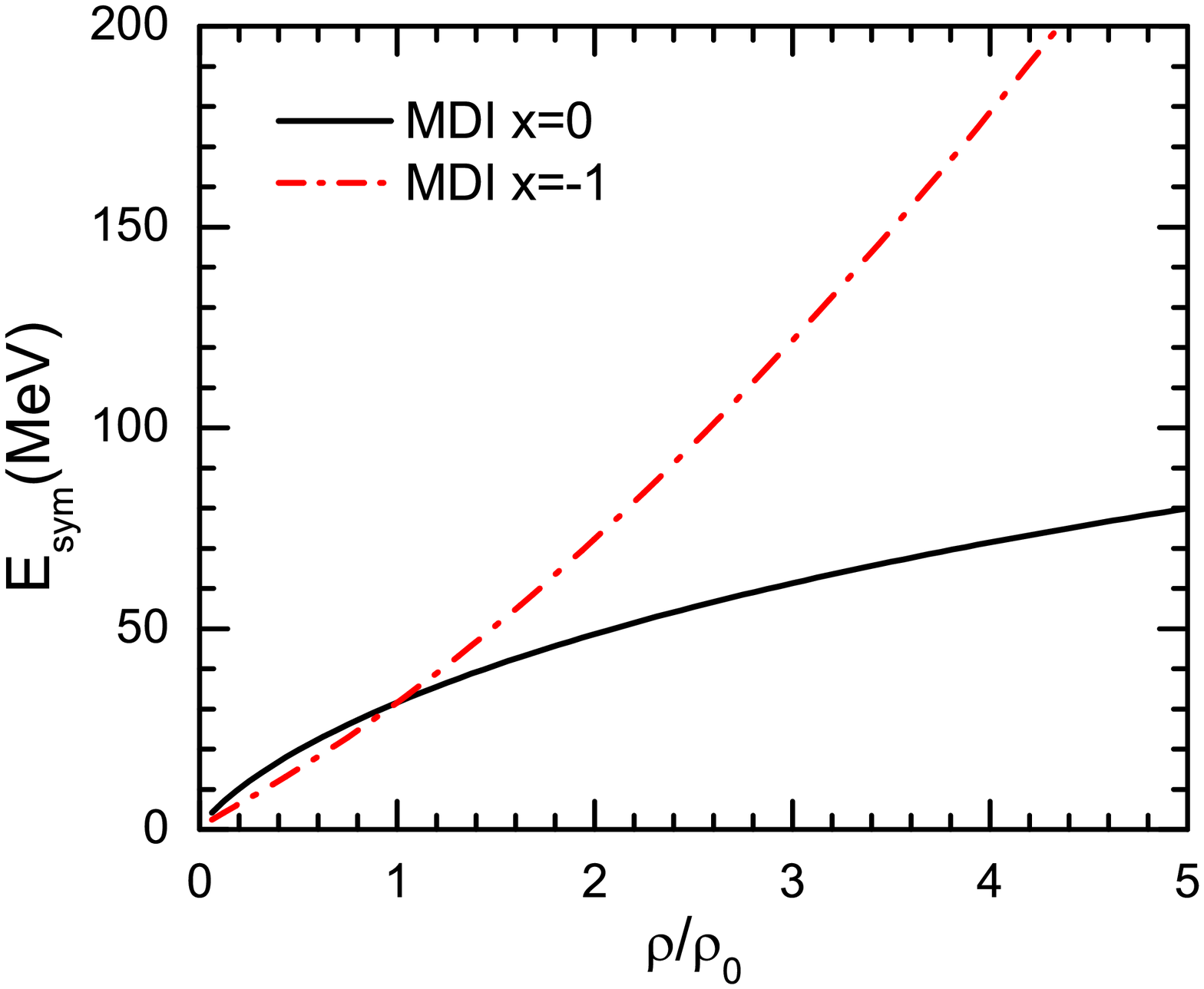}
\end{minipage}
\begin{minipage}{22pc}
\includegraphics[height=4.2cm,width=4.4cm]{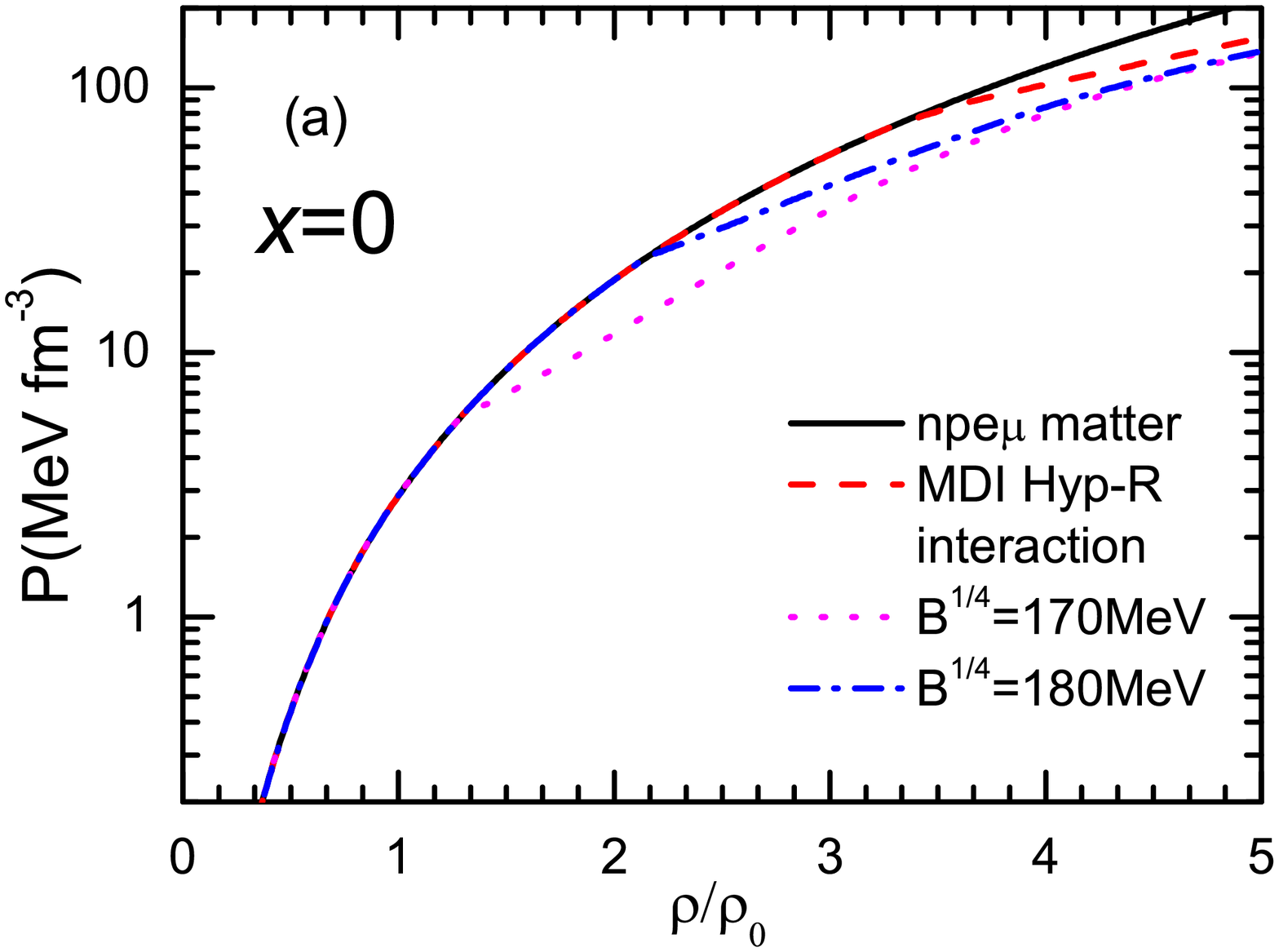}
\includegraphics[height=4.2cm,width=4.4cm]{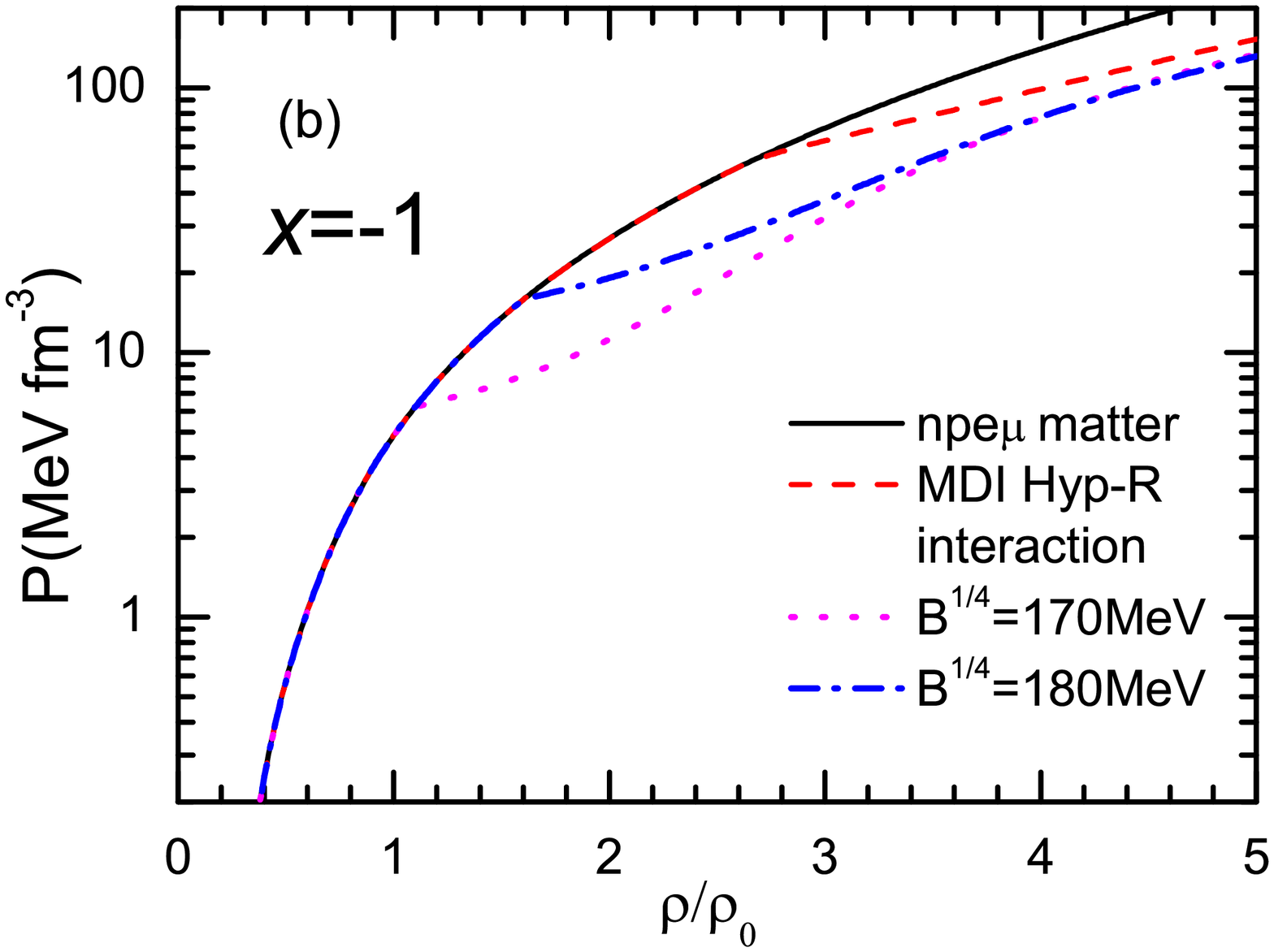}
\end{minipage}
\caption{\textbf{Left Window}: Density dependence of nuclear symmetry
energy from the MDI interaction with parameter $x=0$ and $-1$.
\textbf{Right Windows}: EOSs of pure $npe\mu$ matter (nucleonic),
hyperonic matter (MDI Hyp-R interaction) and hybrid stars for the
bag constant $B^{1/4}=180$ MeV and $170$ MeV with $x=0$ (a) and
$x=-1$ (b).}\label{Wlin}
\end{figure}

The total energy release $E_g$ during the hadron-quark phase
transition is the difference of binding energy before and after
the micro-collapse. It reduces to the difference in the
gravitational masses for the hadronic (h) and hybrid (q)
configurations of the neutron star as a result of baryon number
conservation, namely, $E_g=M_{G,h}-M_{G,q}$. Thus, the $E_g$
should depend on the EOSs of dense matter either in the hadronic
or quark phase. For the neutron-rich hadronic matter, the most
uncertain part of its EOS is the density dependence of the nuclear
symmetry energy, while for the quark matter within the MIT bag
model, the most uncertain part of its EOS is the bag constant. The
relative effects of the density dependence of the nuclear symmetry
energy and the bag constant on the total energy release due to the
hadron-quark phase transition in NSs were recently examined by Lin
et al. \cite{Wklin11}. Shown in Fig.\ \ref{Wlin} are several model
EOSs for hybrid stars obtained from an isospin- and
momentum-dependent effective interaction (MDI)~\cite{das,Xu10} for
the baryon octet, the MIT bag model for the quark
matter~\cite{MIT1,MIT2} and the Gibbs construction for the
hadron-quark phase transition~\cite{Glen92,Glen01}. For
comparisons, the pure $npe\mu$ (labeled as nucleonic) and
hyperonic (labeled as MDI Hyp-R interaction) EOSs are also
included. As it is well known, the appearance of hyperons and the
hadron-quark phase transition softens the EOS of neutron star
matter. The EOSs with $x=0$ and $x=-1$ are about the same below
and around the saturation density. However, it is interesting to
see that the starting point and the degree of softening due to the
appearance of hyperons are sensitive to the $E_{\rm sym}(\rho)$ at
high densities. Moreover, the $E_{\rm sym}(\rho)$ also affects
appreciably the starting point of the hadron-quark mixed phase,
especially with the larger bag constant. Nevertheless, it is
obvious that the starting point is much more sensitive to the bag
constant $B$ for a given symmetry energy parameter $x$. These
features are consistent with those first noticed by Kutschera et
al. \cite{Kut00}.

\begin{figure}[h]
\begin{center}
\begin{minipage}{16pc}
\includegraphics[width=0.96\textwidth]{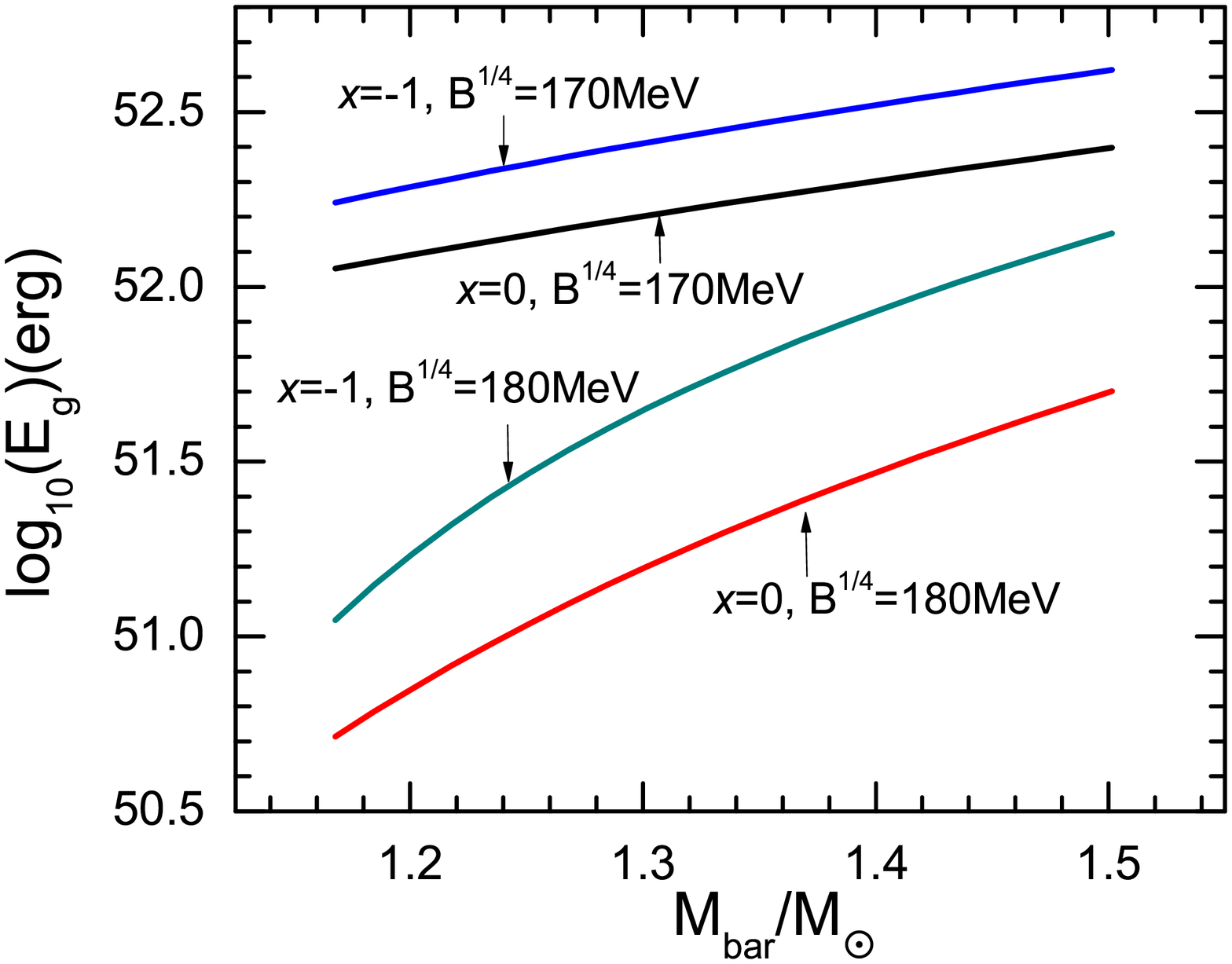}
\end{minipage}
\begin{minipage}{16pc}
\includegraphics[width=1.0\textwidth]{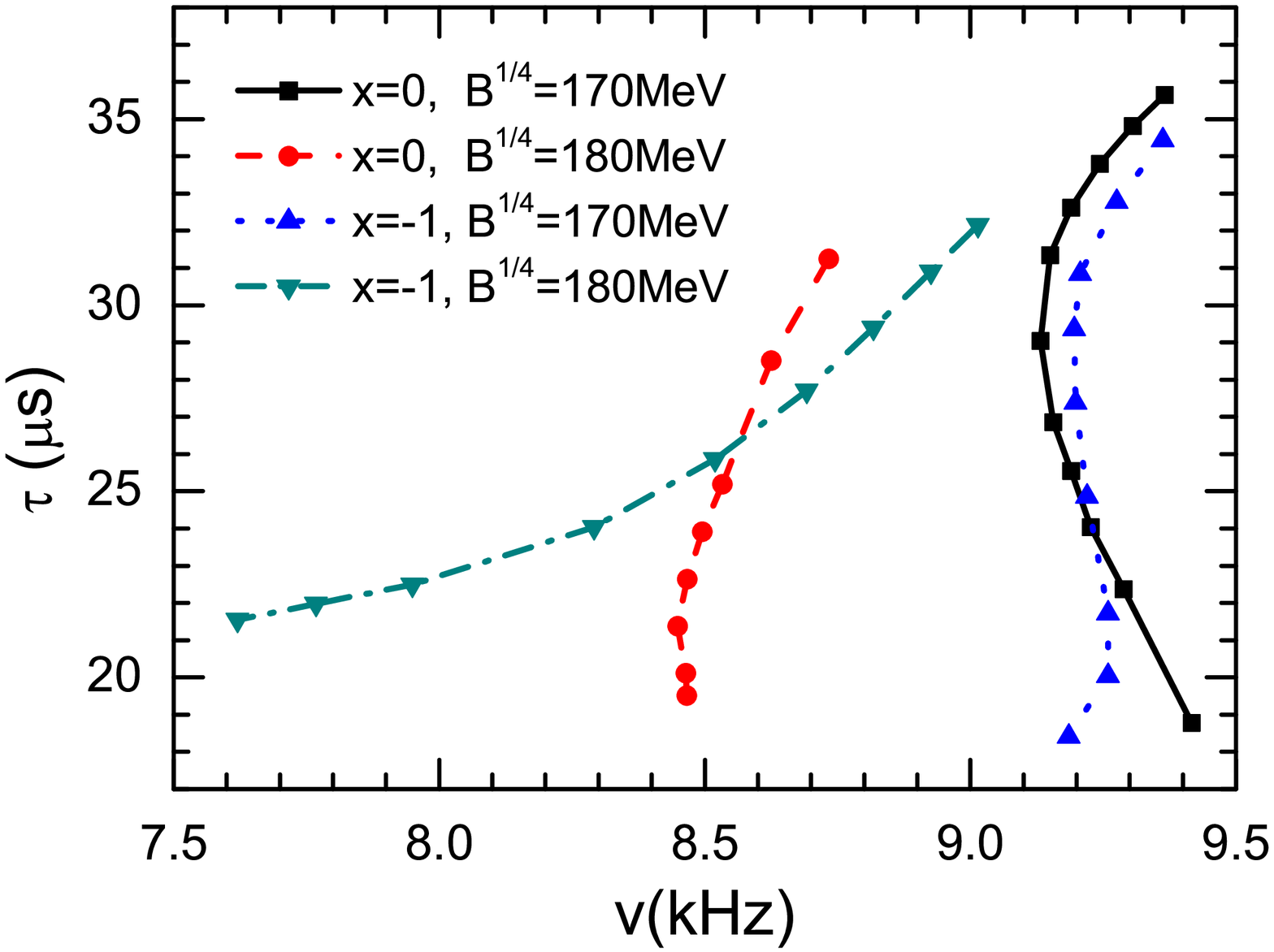}
\end{minipage}
\caption{{\protect\small \textbf{Left Window}: Total energy release
due to the hadron-quark phase transition as a function of the
baryonic mass of a neutron star. \textbf{Right Window}: Damping time
versus frequency of the first $w$-mode for hybrid
stars.}}\label{Ere}\label{tv}
\end{center}
\end{figure}

Shown in the left window of  Fig.\ \ref{Ere} is the energy release
as a function of the baryonic mass $M_{\rm bar}/M_{\odot}$ of a
neutron star. It is seen that the energy release increases with
$M_{\rm bar}/M_{\odot}$ and is higher with a smaller ($B^{1/4}=170$
MeV) bag constant $B$ but a stiffer ($x=-1$) symmetry energy.
Effects of varying the bag constant $B$ are obviously more
significant than varying the symmetry energy parameter $x$,
especially on the core mass and thus the energy release. The
variation of the bag constant $B$ also affects appreciably the radii
of hybrid stars. On the other hand, the variation of the symmetry
energy parameter $x$ only has appreciable effects on the radii of
both hyperonic and hybrid stars. It's effects on the energy release
is much smaller than the bag constant $B$.

\subsection{Imprints of symmetry energy on the w-mode of gravitation waves}

The search for gravitational wave signals is another forefront in
astrophysics and cosmology. Gravitational waves are tiny
disturbances in space-time and are a fundamental, although not yet
directly confirmed, prediction of General Relativity. They would
open up an entirely new non-electromagnetic window for probing the
physics that is hidden or dark to current electromagnetic
observations. There are many possible sources of gravitational
waves. For instance, the hadron-quark phase transition can trigger
various oscillation modes of neutron stars. When coupled with
rotation, these oscillations generate gravitational waves. The
energy release discussed in the previous subsection determines the
maximum strain amplitude of emitted gravitational waves. Moreover,
the strain amplitude of gravitational waves from pulsars depends
on their quadrupole deformation which, in turn, is determined by
the EOS of neutron-rich nucleonic matter. Furthermore, according
to Chandrasekhar \& Ferrari \cite{wmode}, the frequency and
damping time of the $w$-mode of gravitational waves from
oscillating neutron stars are described by a unique second-order
differential equation with the EOS as the only input. Indeed, it
was recently shown that the strain amplitude from elliptically
deformed pulsars as well as the frequency and damping time of the
$w$-mode of oscillating neutron stars are all strongly influenced
by the density dependence of the nuclear symmetry energy
\cite{Kra08b,Wor09,Plamen1,Wen,Plamen2}. Thus, an accurate
determination of the nuclear symmetry energy is also important for
detecting gravitational waves using various ground-based and
satellite-based detectors, such as, LIGO \cite{LIGO} and LISA
\cite{LISA}. As a quantitative example, shown in the right window
of Fig.\ \ref{tv} is the frequency versus damping time of the
first axial $w$-mode. For the softer EOS with $B^{1/4}=170$ MeV,
the effect of the symmetry energy is small. The density dependence
of the symmetry energy can, however, significantly affect the
frequency of the $w$-mode for the stiffer EOS with $B^{1/4}=180$
MeV as a result of the higher hadron-quark transition density. In
this case, the symmetry energy in the hadronic phase then also
plays an important role in determining the structure of NSs.
Nevertheless, comparing the frequencies of the $w$-mode for the
same $x$ parameter but different values of $B$, it is clear that
the bag constant has a much stronger effect as it changes the
underlying EOS of dense matter more significantly.

\section{Summary}

In summary, important progress has been made in recent years in
constraining the symmetry energy, especially around and below the
saturation density, with terrestrial nuclear laboratory data.
Nevertheless, the field is still at its beginning. While a number
of potentially useful probes of the symmetry energy, especially at
supra-saturation densities, have been proposed, available
experimental data are mostly for reactions with stable beams.
Coming experiments with more neutron-rich nuclei at several
advanced high energy radioactive beam facilities are expected to
improve the situation dramatically. Thus, more exciting times are
yet to come. Implications of the partially constrained symmetry
energy on some astrophysical phenomena have also been explored. On
the other hand, impressive progress has also been made in
extracting information about the EOS from astrophysical
observations. Hopefully, working together and combining results
from both fields will allow us to finally pin down the EOS of
neutron-rich nuclear matter at both sub- and super-saturation
densities in the near future.

\section{Acknowledgments} This work was supported in part by the
US National Science Foundation Grant Nos. PHY-0757839 and
PHY-0758115, the National Aeronautics and Space Administration under
grant NNX11AC41G issued through the Science Mission Directorate, the
Welch Foundation under Grant No.\ A-1358, the Texas Coordinating
Board of Higher Education Grant No.003565-0004-2007, the Young
Teachers' Training Program from China Scholarship Council under
Grant No. 2007109651, the National Natural Science Foundation of
China under Grant No.10947023, 10735010,10775068,10805026 and
10975097, and the Fundamental Research Funds for the Central
University, China under Grant No.2009ZM0193, Shanghai Rising-Star
Program under grant No. 11QH1401100, the National Basic Research
Program of China (2007CB815004 and 2010CB833000).

\end{document}